\documentclass[12pt,english]{iopart}
\usepackage[T1]{fontenc}
\usepackage[latin1]{inputenc}
\usepackage{graphicx}

\makeatletter
\usepackage{graphicx}
\usepackage{iopams}

\usepackage{babel}
\makeatother
\begin{document}
\title{Accelerating random walks by disorder}
\author{Vitaly Belik$^1$ and Dirk Brockmann}
\address{Max Planck Institute for Dynamics and Self-Organization, 37073, Bunsenstr. 10, G\"ottingen, Germany}
\ead{$^1$ vitaly@nld.ds.mpg.de}
\begin{abstract}
We investigate the dynamic impact of heterogeneous environments on superdiffusive random walks known as L\'evy flights. We devote particular attention to the relative weight of source and target locations on the rates for spatial displacements of the random walk. Unlike ordinary random walks which are slowed down for all values of the relative weight of source and target, non-local superdiffusive processes show distinct regimes of attenuation and acceleration for increased source and target weight, respectively. Consequently, spatial inhomogeneities can facilitate the spread of superdiffusive processes, in contrast to common belief that external disorder generally slows down stochastic processes. Our results are based on a novel type of fractional Fokker-Planck equation which we investigate numerically and by perturbation theory for weak disorder.
\end{abstract}
\submitto{New Journal of Physics}

A particle performing ordinary diffusion is typically characterized
by spatiotemporal scaling relation $|\bi{X}(t)|\sim t^{1/2}$. An
increasing number of physical and biological systems are in conflict
with this relation and exhibit anomalous diffusion. Whenever $|\bi{X}(t)|\sim t^{1/\mu}$
with an exponent $0<\mu<2$ a system is said to exhibit superdiffusive
behavior. Superdiffusion was discovered in a wide range of systems,
for instance chaotic dynamical systems \cite{Geisel}, particles in
turbulent flows \cite{Bodenschatz}, saccadic eye movements~\cite{DBeyes},
trajectories of foraging animals~\cite{albatros}, and very recently
in the geographic dispersal of bank notes~\cite{DBNature}. 

Theoretically, superdiffusion is often accounted for by scale free
random walks known as L\'evy flights for which successive spatial displacements
$\Delta\mathbf{x}$ are drawn from a probability density function
(pdf) with an algebraic tail, i.e. $p(\Delta\bi{x})\sim|\Delta\bi{x}|^{-d-\mu}$,
where $d$ is the spatial dimension and the L\'evy exponent $\mu<2$,
such that the variance of $\Delta\bi{x}$ is divergent~\cite{SHLES}.
Embedded in the context of continuous time random walks (CTRW) one
can show that the dynamics of the pdf of finding a superdiffusive
particle at position $\bi{x}$ at time $t$ is governed by a fractional
generalization of the diffusion equation,\begin{equation}
\partial_{t}p(\bi{x},t)=D_{\mu}\Delta^{\mu/2}p(\bi{x},t),\label{eq:fracdiff}\end{equation}
in which the ordinary Laplacian $\Delta$ is replaced by the fractional
operator $\Delta^{\mu/2}$, a non-local integral operator, the action
of which is equivalent to a multiplication by $-|\bi{k}|^{\mu}$ in
Fourier space. The constant $D_{\mu}$ is the generalized diffusion
coefficient.

A process governed by Eq.~(\ref{eq:fracdiff}) is spatially homogeneous
and isotropic, as the probability rate $w(\bi{y}|\bi{x})$ of a displacement
from $\bi{x}$ to $\bi{y}$ depends only on distance, i.e. $w(\bi{y}|\bi{x})\propto|\bi{x}-\bi{y}|^{-(d+\mu)}$.

Numerous random processes, however, occur on spatially disordered
substrates or evolve in the presence of quenched spatial inhomogeneities.
Depending on the underlying physical or biological system, one obtains
various generalizations of ~(\ref{eq:fracdiff}) which incorporate
the spatial structure. In the generalized Langevin approach~\cite{Fogedby,MK,MK2}
an additional force term $-\nabla F$ on the right hand side of ~(\ref{eq:fracdiff})
accounts for an external position dependent force field $F(\mathbf{x})$.
In topologically superdiffusive systems, such as intersegment transfer
of gene regulatory enzymes on DNA strands~\cite{DBPRL2,Metzler,Lomholt},
the transition rate is modified by a Boltzmann factor, i.e. $w(\bi{y}|\bi{x})\propto|\bi{x}-\bi{y}|^{-(d+\mu)}\times\exp-\beta[V(\bi{y})-V(\bi{x})]/2$,
where $V(\bi{x})$ is a position dependent potential and $\beta$
an inverse temperature. In subordinated superdiffusive processes,
an ordinary diffusion process subjected to an external force is sampled
at highly variable operational time intervals~ \cite{SUB}. All three
systems exhibit very different response properties to the imposed
spatial structure~\cite{BS}, but converge to the same Fokker-Planck
equation in the limit of ordinary diffusion. 

\begin{figure}
\includegraphics[scale=1.2]{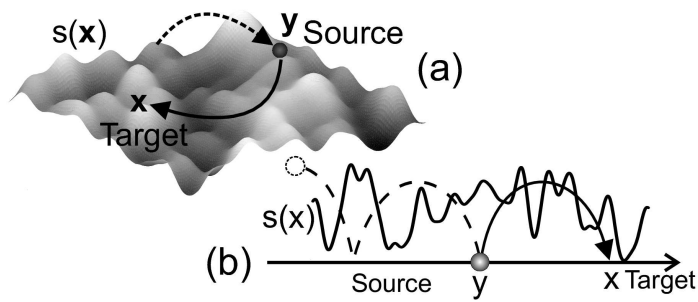}
\caption{Random walk processes in inhomogeneous salience fields $s(\mathbf{x})$
in two (a) and one (b) dimensions. Source and target locations of
a random jump are denoted by $\mathbf{y}$ and $\mathbf{x}$, respectively.}

\label{sal}
\end{figure}

Here we investigate the dynamic impact of heterogeneous environments
and devote particular attention to the relative impact of source and
target locations on the rates $w(\bi{x}|\bi{y})$ for spatial displacements
of the random walk (see Fig.\ref{sal}). We define the spatial inhomogeneity
in terms of the attractivity or salience $s(\bi{x})>0$ of a location
$\bi{x}$. For large and small values of $s(\bi{x})$, the location
$\bi{x}$ is attractive and unattractive, respectively. We assume
that in equilibrium a walker's stationary probability $p^{\star}(\bi{x})$
of being at a location $\bi{x}$ is proportional to the salience at
$\bi{x}$, i.e. $p^{\star}(\bi{x})\propto s(\bi{x})$. In that respect,
the salience field can be defined operationally as the likelyhood
of finding a walker at site $\bi{x}$. Furthermore we assume that
a transition from $\bi{y}$ to $\bi{x}$ is more likely to occur when
the salience is large at the target location $\bi{x}$, and less likely
to occur when the salience is large at the source location $\bi{y}$.
This leads to a transition rate\begin{equation}
w(\bi{x}|\bi{y})=\frac{1}{\tau}s^{c}(\bi{x})f(|\bi{x}-\bi{y}|)s^{c-1}(\bi{y}),\label{TR}\end{equation}
where $\tau$ is a time constant and the sandwiched term $f(|\bi{x}-\bi{y}|)\propto|x-y|^{-(d+\mu)}$
with $0<\mu\leq2$ accounts for L\'evy flight jump lengths. Inserting
~(\ref{TR}) into the master equation\begin{equation}
\partial_{t}p(\bi{x},t)=\int\rmd\bi{y}\, w(\bi{x}|\bi{y})p(\bi{y},t)-w(\bi{y}|\bi{x})p(\bi{y},t)\label{eq:master}\end{equation}
one can see that detailed balance is fullfilled and that the stationary
solution if it exists is proportional to the salience. The central
parameter in our analysis is the weight parameter $0\leq c\leq1$,
which quantifies the relative impact of source and target salience
on the dynamics. When $c=1$ a transition $\bi{y}\rightarrow\bi{x}$
only depends on the salience at the target site and is independent
of the salience at the source, when $c=0$ the salience of the target
site has no influence on the transition and the rate is decreased
with increasing salience at the source. The intermediate case $c=1/2$
is equivalent to topological superdiffusion approach with a salience
given by a Boltzmann factor $s(\bi{x})=e^{-\beta V(\bi{x})}$. 

In combination with the rate (\ref{TR}) the master ~(\ref{eq:master})
is equivalent to the fractional Fokker-Planck equation\begin{equation}
\partial_{t}p=D\, s^{c}\Delta^{\mu/2}s^{(c-1)}p-D\, p\, s^{c-1}\Delta^{\mu/2}s^{c},\label{OE}\end{equation}
where $p=p(\bi{x},t)$, $s=s(\bi{x})$ and $D$ is a generalized diffusion
coefficient. The fractional Laplacian is defined by \[
\Delta^{\mu/2}f(\bi{x})=\mathcal{C}_{\mu}\int\rmd\bi{y}\frac{f(\bi{y})-f(\bi{x})}{|\bi{x}-\bi{y}|^{d+\mu}}\]
with $\mathcal{C}_{\mu}=2^{\mu}\pi^{-n/2}\Gamma((\mu+n)/2)/\Gamma(-\mu/2)$.
In Fourier space $\Delta^{\mu/2}$ corresponds to multiplication by
$-|\bi{k}|^{\mu}$: $\mathcal{F}\{\Delta^{\mu/2}f\}(\bi{k})=-|\bi{k}|^{\mu}\mathcal{F}\{ f\}(\bi{k})$.
Note that the fractional Fokker-Planck ~(\ref{OE}) is equivalent
to a number of known stochastic processes for specific choices of
the parameters $c$ and $\mu$. For instance when $s(\bi{k})={\rm const}$
~(\ref{OE}) is equivalent to free superdiffusion, i.e. ~(\ref{eq:fracdiff}),
when $c=1/2$ and $\mu=2$, (\ref{OE}) reads $\partial_{t}p=-\nabla F\, p+\Delta p$,
with $F(\bi{x})=-V^{\prime}(\bi{x})$ and $V(\bi{x})=-\beta^{-1}\log s(\bi{x})$,
i.e. diffusion in an external force field. When $c=0$, (\ref{OE})
reads $\partial_{t}p=D\Delta^{\mu/2}p/s$, which is equivalent to
generalized multiplicative Langevin dynamics for the process $\bi{X}(t)$,
i.e. $\rmd\bi{X}=D(\bi{X})\rmd L_{\mu}$ with $D(\bi{X})=\sqrt{2}(\exp\beta V(\bi{X})/2)/D$
and $L_{\mu}(t)$ is a homogeneous L\'evy stable process.

In the following we investigate the relaxation properties of one dimensional
processes governed by the fractional Fokker-Planck equation~(\ref{OE}).
It is convenient to make a transformation of variables $\psi=s^{1/2}p$
and rewrite the dynamics as a generalized Schr�inger equation $\partial_{t}\psi=\mathcal{H}\psi$
with a Hamiltonian \begin{equation}
\mathcal{H}\psi=s^{c-1/2}\Delta^{\mu/2}s^{c-1/2}\psi-\psi s^{c-1}\Delta^{\mu/2}s^{c}\label{eq:SHR}\end{equation}
possessing identical spectral properties.

We model inhomogeneities as salience fields of the type $s=\exp(-\varepsilon v)$,
where $v=v(x)$ is a random potential with zero mean and unit variance.
The parameter $\varepsilon>0$ quantifies the strength of the inhomogeneity.
Typically, inhomogeneities posses a spatial correlation length. For
$v$ we chose random phase potentials defined by 

\begin{equation}
v(x)=\frac{1}{2\pi}\int\rmd k\phi(k)e^{i\theta(k)-ikx}.\label{rpp}\end{equation}
 Here the phase $\theta(k)$ is uniformly distributed on the interval
$(0,2\pi]$ and the amplitude is given by the correlation spectrum
$S(k)$, $\phi(k)\overline{\phi(k')}=2\pi S(k)\delta(k-k')$, which
we chose to be Gaussian with a correlation length $\xi$: $S(k)=2\xi\exp(-k^{2}\xi^{2}/\pi)$.
For small $\varepsilon$ we can expand (\ref{eq:SHR}) up to the second
order and obtain a Hamiltonian of the form $\mathcal{H}=\Delta^{\mu/2}+\hat{U}$
which we can treat perturbation theoretically to obtain a spectrum
of the form $E(k)\approx-D_{\mu,c}(k;\varepsilon)|k|^{\mu}$. Here
$D_{\mu,c}(k;\varepsilon)$ describes the relaxation properties on
the corresponding scale length $\lambda=k^{-1}$. In the limit of
vanishing potential the generalized diffusion coefficient is identical
to the diffusion coefficient $D$ of the free system. Typically, a
random spatial inhomogeneity slows down a random process, and one
generally expects $D_{\mu,c}(k;\varepsilon)$ to be smaller than $D_{\mu,c}(k;\varepsilon=0)$.
Up to the second order in $\varepsilon$ we obtain \begin{equation}
D_{\mu,c}(k;\varepsilon)/D=1-\varepsilon^{2}G_{\mu,c}(k),\label{1b2}\end{equation}
 where $G_{\mu,c}(k)=\frac{1}{2\pi}\int\rmd qS(q)g_{\mu,c}(k/q)$
and\begin{eqnarray}
g_{\mu,c}(z) & = & -2(c-1/2)^{2}+\frac{1}{z^{\mu}}\Big\{2c(c-1)-(c-1/2)^{2}(|1-z|^{\mu}+|1+z|^{\mu})\nonumber \\
 & + & [(c-1/2)(z^{\mu}+|1+z|^{\mu})-c]^{2}/(|1+z|^{\mu}-z^{\mu})\nonumber \\
 & + & [(c-1/2)(z^{\mu}+|1-z|^{\mu})-c]^{2}/(|1-z|^{\mu}-z^{\mu})\Big\}\label{gsmall}\end{eqnarray}

\begin{figure}
\includegraphics[scale=1.2]{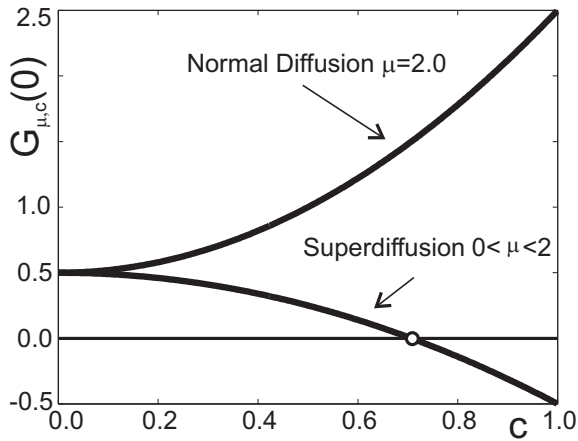}
\caption{The impact of source and target location on the asymtotics of random
walk processes. The generalized diffusion coefficient depends quadratically
on the magnitude $\varepsilon$of the inhomogeneity: $D_{\mu,c}=1-G_{\mu,c}(0)\varepsilon^{2}$.
The pre-factor $G_{\mu,c}(0)$ quantifies the impact of the spatial
inhomogeneity, i.e. when $G_{\mu,c}(0)>0$ the process is slowed down,
when $G_{\mu,c}(0)<0$ the process is accelerated. The two lines depict
$G_{\mu,c}(0)$ as a function of the weight parameter $c$ for ordinary
diffusion ($\mu=2$, upper line) and L\'evy flights ($\mu<2$, lower
line). As c is increased (increasing weight of the target location),
ordinary diffusion is slowed down more strongly contrary to L\'evy flights
for which attenuation decreases until a critical value of $c_{{\rm crit}}=1/\sqrt{2}$
is reached (denoted by small circle). Beyond this point L\'evy flights
are accelerated by the external inhomogeneity.\label{G}}
\end{figure}

The limit $k\rightarrow0$ yields the long scale asymptotics for the
process for which we obtain\begin{equation}
G_{\mu,c}(0)=\left\{ {1/2-c^{2},\,0<\mu<2,\atop 1/2+2c^{2},\,\mu=2,}\right.\label{Glim}\end{equation}
This result indicates that with increasing influence of the target
salience (increasing values of $c$) superdiffusive processes ($\mu<2$)
exhibit the opposite behavior as ordinary diffusion processes ($\mu=2$).
As $c$ is increased $G_{\mu,c}(0)$ increases as well for ordinary
diffusion, which means that these processes are attenuated more strongly.
Quite contrary to superdiffusive processes, for which a more pronounced
target influence decreases $G_{\mu,c}(0)$. This implies that slowing
down of superdiffusive processes becomes weaker as the target weight
in the rates is increased. Note also that for weak potentials the
magnitude of this acceleration is independent of the L\'evy exponent
$\mu$, see Fig.\ref{G}. Only when $c=0$, and target salience has
no impact on the transition rate, all processes are slowed down by
the same amount, i.e. $G_{\mu,0}(0)=1/2$. 

Note that, as $c$ is increased for $\mu<2$, the function $G_{\mu,c}(0)$
even becomes negative for $c>c_{{\rm crit}}=1/\sqrt{2}$. This implies
that these processes are no longer slowed down by the spatial inhomogeneity
but rather accelerated, as a negative value for $G_{\mu,c}(0)$ implies
a diffusion coefficient larger than the one for free superdiffusion,
i.e. $D_{\mu,c}(k\rightarrow0;\varepsilon)>D$. Consequently, the
common notion that random processes are typically slowed down by spatial
inhomogeneities is not valid when superdiffusive processes are involved.

\begin{figure}
\includegraphics[scale=1.2]{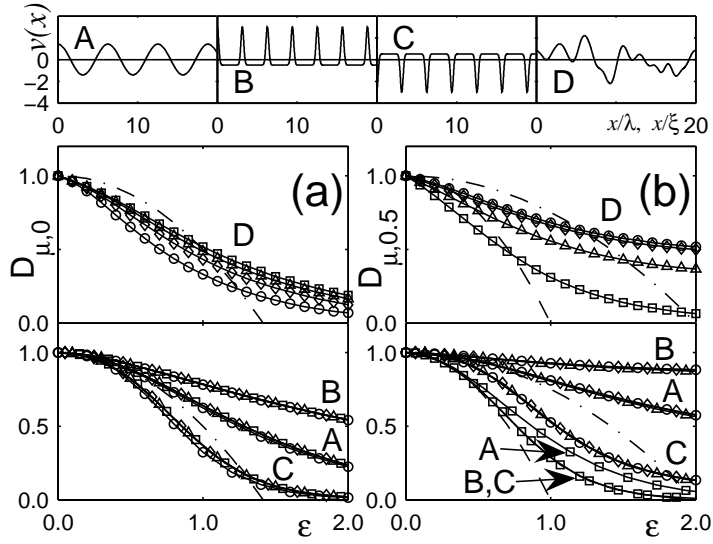}
\caption{Generalized diffusion coefficient $D_{\mu,c}$ as a function of the
magnitude $\varepsilon$ of the salience field $s(x)=\exp[-\varepsilon v(x)]$.
The function $D_{\mu,c}$ is shown for four types of potentials, a
cosine potential (A), potentials with localized maxima and minima,
(B and C, respectively) and a random phase potential (D). Potentials
are displayed on the top panel. Results for two values of the weight
parameter $c$ are presented on the panel (a) $c=0$ and (b) $c=0.5$
correspondingly. Dashed line and dash-dotted lines are analytical
results for $\mu=2$ and $0<\mu<2$ we obtained from perturbation
theory (\ref{Glim}). Each type of symbol corresponds to the one L\'evy
exponent: $\opencircle$- $\mu=0.5$, $\opendiamond$ - $\mu=1.0$,
$\opentriangle$ - $\mu=1.5$, $\opensquare$ - $\mu=2.0$. \label{DK}}
\end{figure}

For large $\varepsilon$ we computed the generalized diffusion coefficient
$D_{\mu,c}(\varepsilon)$ numerically for three periodic and one random
phase potential: a cosine potential $v(x)=\sqrt{2}\cos(x/\lambda)$,
a potential with localized potential minima and one with localized
potential peaks, for which $v(x)=\pm a\cos^{\gamma}(x/\lambda)+b$
($a,b>0,\,\gamma=32$), and a potential defined by  (\ref{rpp}) with
a gaussian spectrum. The results are depicted in Fig.\ref{DK}.

\begin{figure}
\includegraphics[scale=1.2]{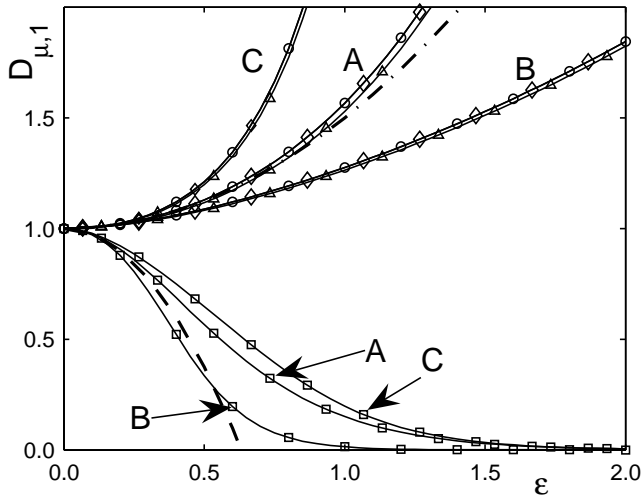}
\caption{Generalized diffusion coefficient $D_{\mu,c}$ as a function of the
magnitude $\varepsilon$ of the different salience fields for the
case $c=1$. The meaning of the symbols see caption to Fig.\ref{DK}.
In contrast to diffusive processes, which are attenuated, the superdiffusive
ones are enhanced by inhomogeneity.\label{DKc1}}
\end{figure}

For $c=0$ (i.e. full weight of source location) the processes in
a given environment behave similarly, independent of the exponent
$\mu$ (Fig.\ref{DK}(a)), ordinary diffusion and all superdiffusive
processes exhibit the same quadratic decrease of $D_{\mu,c}$ with
$\varepsilon$ in a fixed potential. For $c=1/2$ (Fig.\ref{DK}(b))
only the superdiffusive processes exhibit an identical response to
a given potential, the response of the ordinary diffusion process
differs. When $c$ is increased beyond the critical value the difference
between ordinary diffusion and L\'evy flights becomes maximal and changes
qualitatively (Fig.\ref{DKc1}). In this regime ordinary diffusion
processes are still slowed down by the spatial inhomogeneity as opposed
to the L\'evy flights which exhibit an increase in the diffusion coefficient
with increasing potential strength and are thus accelerated. For small
values of $\varepsilon$ the numerics agree well with our results
obtained by the perturbation theory, above (i.e. (\ref{Glim})). Despite
the fact, that for all values of $\mu<2$, the curves collapse on
one single function, L\'evy flights are sensitive to the potential shape.
The relatively pure performance of the random phase potential (D on
the Fig.\ref{DK}) is due to the finite size effects, which does not
constitute an obstacle for the periodical potentials (A,B,C). Therefore
we have omitted the results for the random phase potential in case
$c=1$ (Fig. \ref{DKc1}).

\begin{figure}
\includegraphics[scale=1.2]{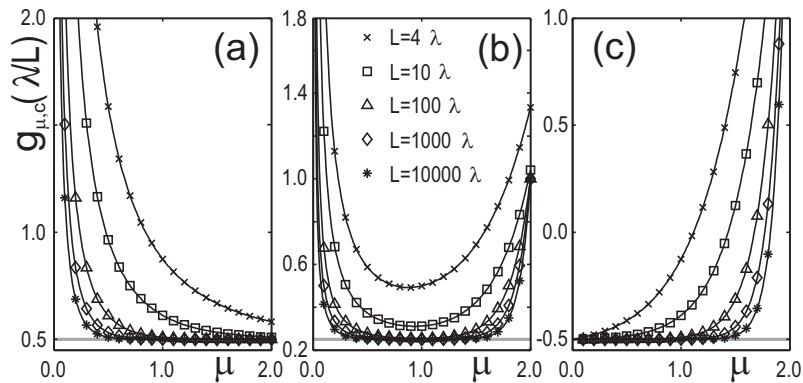}
\caption{The impact of the heterogeneity on the relaxation time of the processes
$\delta t\sim g_{\mu,c}(\lambda/L)$ with different power-law exponents
$\mu$, in dependence on the size of the system $L$ for the cosine
potential with period $\lambda$. Gray lines denote the large scale
limit, according to (\ref{Glim}).}

\label{Bounded}
\end{figure}

The above considerations were restricted solely to the infinite systems.
A key question is how these processes behave in finite systems and
to what extent finite size effects play a role. Therefore we investigate
the relaxation properties in finite systems of length $L$. To this
end we consider the quantity \[
\delta\tau=\tau/\tau_{{\rm free}}-1,\]
where $\tau$ is the relaxation time of the process and $\tau_{{\rm free}}$
is the relaxation time of the same process without spatial inhomogeneity.
For the cosine potential we obtain $\delta\tau\approx\varepsilon^{2}g_{\mu,c}(\lambda/L)$,
where $g_{\mu,c}$ is defined in (\ref{gsmall}). See Fig.\ref{Bounded}
for three values of $c=0,\,0.5$ and $1$. Small values of $\delta\tau$
correspond to a small effect of the inhomogeneity. For $c=0$ (Fig.~\ref{Bounded}(a)),
the ordinary diffusion process exhibits the smallest value of $\delta\tau$,
which implies that in situations in which the source salience is important,
diffusion relaxes fastest. On the contrary, for $c=1$ (Fig.~\ref{Bounded}(c))
strongly superdiffusive processes (i.e. $\mu\rightarrow0$) exhibit
a small $\delta\tau$. Only when both, source and target possess an
equal impact on the jump rates (i.e. $c=0.5$ (\ref{Bounded}(b)),
$\delta\tau$ exhibits a minimum for intermediate values of the L\'evy
exponent $\mu$~\cite{DBPRL}. %

We considered the consequences of the relative weight of source and
target locations in one-dimensional random walk processes, evolving
in inhomogeneous environment. Our analysis revealed essential differences
between superdiffusive L\'evy flights and ordinary random walks when
they occur in regular and random spatial inhomogeneities. Unlike ordinary
random walks, L\'evy flights can be accelerated when the influence of
the target salience is sufficiently large, which may shed a new light
on optimal search strategies in heterogeneous landscapes, and dispersal
phenomena in population dynamical systems and various physical and
biological contexts.

\ack We thank O.A. Chichigina, L. Hufnagel, T. Geisel, W. Noyes,
and I.M. Sokolov for their support, valuable discussions and comments.

\section*{References}

\bibliographystyle{unsrt}
\bibliography{iopref}

\begin{thebibliography}{10}

\bibitem{Geisel}
Geisel T.
\newblock {\em Phys. Rev. Lett.}, 54:616, 1985.

\bibitem{Bodenschatz}
Porta~A L, Voth~G A, Crawford~A M, Alexander J, and Bodenschatz E.
\newblock {\em Nature (London)}, 409:1017, 2001.

\bibitem{DBeyes}
Brockmann D and Geisel T.
\newblock {\em Neurocomputing}, 32-33:643, 2000.

\bibitem{albatros}
Viswanathan~G M, Buldyrev~S V, Havlin S, da~Luz M E~G, Raposo~E P, and
  Stanley~H E.
\newblock {\em Nature (London)}, 401:911, 1999.

\bibitem{DBNature}
Brockmann D, Hufnagel L, and Geisel T.
\newblock {\em Nature (London)}, 439:462, 2006.

\bibitem{SHLES}
Shlesinger~M F and Zaslavsky~G M.
\newblock Lecture notes in physics. Springer-Verlag, Berlin, 1995.

\bibitem{Fogedby}
Fogedby~H C.
\newblock {\em Phys. Rev. Lett. E}, 58:1690, 1998.

\bibitem{MK}
Metzler R and Klafter J.
\newblock {\em Phys. Rep.}, 339:1, 2000.

\bibitem{MK2}
Metzler R and Klafter J.
\newblock {\em J.Phys. A: Math. Gen.}, 37:R161, 2004.

\bibitem{DBPRL2}
Brockmann D and Geisel T.
\newblock {\em Phys. Rev. Lett.}, 91:048303, 2003.

\bibitem{Metzler}
Sokolov~I M, Metzler R, Pant K, and Williams~M C.
\newblock {\em Biophys. J.}, 89:895, 2005.

\bibitem{Lomholt}
Lomholt~M A, Ambjornsson T, and Metzler R.
\newblock {\em Phys. Rev. Lett.}, 95:260603, 2005.

\bibitem{SUB}
Sokolov~I M.
\newblock {\em Phys. Rev. E}, 63:011104, 2000.

\bibitem{BS}
Brockmann D and Sokolov~I M.
\newblock {\em Chem. Phys.}, 284:409, 2002.

\bibitem{DBPRL}
Brockmann D and Geisel T.
\newblock {\em Phys. Rev. Lett.}, 90:170601, 2003.

\end{thebibliography}

\end{document}